\documentstyle[11pt,psfig]{article}
\def\tts{\small\tt }

\begin{document}
\vspace*{-17mm}
\hfill Data Science Journal, Volume 8 (2009), in press  (http://dsj.codataweb.org) \\[1mm]

{\large\bf SAFEGUARDING OLD AND NEW JOURNAL TABLES FOR THE VO: 
STATUS FOR EXTRAGALACTIC AND RADIO DATA}

\textrm{\bfseries\slshape Heinz Andernach}

{\it Argelander-Institut f\"ur Astronomie, Universit\"at Bonn, D-53121 Bonn, Germany; \\
(on leave of absence from Depto.\ de Astronom\1a, Universidad de Guanajuato, Mexico) \\
Email:} \underline{heinz@astro.ugto.mx} \\[2ex]

\centerline{\textrm{\bfseries\slshape ABSTRACT}}

{\it
Independent of established data centers, and partly for my own research,
since 1989 I have been collecting the tabular data from over 2600 articles
concerned with radio sources and extragalactic objects in general. Optical
character recognition (OCR) was used to recover tables from 740
papers. Tables from only 41 percent of the 2600 articles are available in
the CDS or CATS catalog collections, and only slightly better coverage is
estimated for the NED database. This fraction is not better for articles
published electronically since 2001. Both object databases (NED, SIMBAD,
LEDA) as well as catalog browsers (VizieR, CATS) need to be consulted to
obtain the most complete information on astronomical objects. More human
resources at the data centers and better collaboration between authors,
referees, editors, publishers, and data centers are required to improve
data coverage and accessibility. The current efforts within the Virtual
Observatory (VO) project, to provide retrieval and analysis tools for
different types of published and archival data stored at various sites,
should be balanced by an equal effort to recover and include large
amounts of published data not currently available in this way.
}


\bigskip
{\large\bf 1~~~~~~INTRODUCTION}

During the 1980s, the amount of information on astronomical sources
had grown to such an extent that it was necessary to create databases
that would not only provide links between the different designations
of a certain object in different wavebands of the electromagnetic
spectrum, from radio through X-rays, but would also include a complete
bibliography of each object.  As a result, SIMBAD was created at the
Centre de Données Strasbourg (CDS), France (Wenger, Ochsenbein, Egret,
Dubois, Bonnarel, Borde, et al.\ 2000), initially for stars in our Galaxy
and later expanded to include extragalactic objects, as well as LEDA at
Observatoire de Lyon, France (Paturel, Andernach, Bottinelli, Di Nella,
Durand, Garnier R., et al.\ 1997) for galaxies of the nearby Universe,
and eventually NED at IPAC, USA (Mazzarella \& The NED Team, 2007) to
include all extragalactic objects known.  In parallel to these databases,
CDS Strasbourg maintains a growing collection of astronomical catalogs
in electronic form, which are individually accessible for download and
have become searchable through the VizieR catalog browser (Ochsenbein,
Bauer, \& Marcout, 2000; see URL vizier.u-strasbg.fr/cgi-bin/VizieR).

In 1989, motivated by a lack of data on radio continuum sources in NED,
SIMBAD, and VizieR, I started collecting electronic tables of radio
sources and/or extragalactic objects that were not readily available
from data centres (see e.g., Andernach, 1992). This has grown into a
collection of data tables from currently over 2600 articles. The history
and current status of this collection is described in Sect.\,2; while in
Sect.\,3 the CDS catalog collection is analysed. In Sect.\,4 the size
distribution of catalogs in my collection is studied, and in Sect.\,5 
the fraction of the literature that is covered by more established
catalog collections is estimated. In Sect.\,6 the coverage of the purely
electronic literature since 2001 is investigated, and in Sect.\,7 the
most complete database for extragalactic objects is examined for its
catalog coverage. Sect.\,8 presents my conclusions.

\bigskip
{\large\bf 2~~~~~~~GROWTH AND STATUS OF AUTHOR'S CATALOG COLLECTION}

After having started my collection of electronic catalogs not initially
available from data centers in 1989, by 1995 I had secured over 130 radio
source lists. Half of the latter, typically the larger ones, had been
made available for cone searches in the Einstein On-line Service (Harris,
Stern Grant, \& Andernach, 1995) until that service was closed. Since 1997,
a larger number of radio catalogs from my collection were integrated
into the CATS database (Verkhodanov, Trushkin, Andernach, \& Chernenkov
1997) and a smaller amount into the VizieR catalog browser. The growth
of my collection over time, independent of size and publication year,
is plotted in Figure~1. Catalogs were selected if they (a) contained
data on an appreciable number (approximately at least 50) of radio
continuum sources or other extragalactic objects and (b) were initially
unavailable in the CDS catalog collection. Many of these catalogs were
later incorporated in the CDS archive and VizieR, but, as I shall show in
this contribution, more than half of the tables in my collection are still
not accessible from these public catalog browsers. Until the present,
I collected radio source lists from 1380 articles published since the
1950s and other tables including extragalactic objects from a further
1250 articles. About a third of these were taken from the arXiv preprint
server at (URL www.arXiv.org), about another third were recovered with
optical character recognition (OCR) methods, and the rest was either
supplied by authors upon my request (for papers before about 1995)
or recovered from electronically published articles (since about 1995).

\begin{figure}[!b]
\hspace*{23mm}
\psfig{file=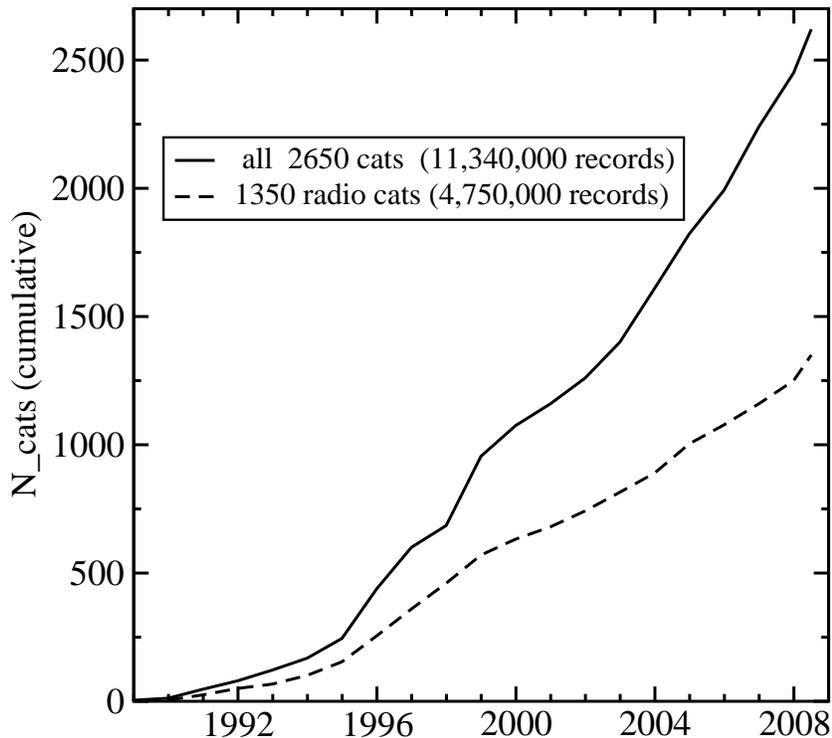,width=11cm,bbllx=549pt,bblly=13pt,bburx=87pt,bbury=525pt,angle=-90,clip=}
\caption{Cumulative growth of the author's catalog collection. For
each year on the horizontal axis, the number of published articles from
which tabular data were collected until that year (not the year of the
publication of each catalog) is plotted on the vertical axis. Continuous
line: all collected catalogs; broken line: radio source lists only.}
\end{figure}

The initial motivation for this collection was to convince the then
established data centers (CDS and NASA's Astronomical Data Center, ADC)
to aim for better literature coverage in their catalog collections. The
hope that my collection work would soon become redundant did not become
a reality. On the contrary, the collection work by established data
centers was further reduced with the closure of NASA's ADC in 2002.
As Figure~1 shows, my collection has been growing at an increasing
speed over the years and is currently growing at 200 items per year
where I define an ``item'' as one article with at least one table in
my collection. In Figures 1 through 4, these items are also denoted as
``cats'' (short for catalogs).

Given the incompleteness of the catalog collections of CDS and CATS,
the author is continuing to collect newly published catalogs if they are
unlikely to enter these collections and to recover older catalogs which
have never existed in electronic form by using OCR methods. To this end,
the OCR results of the scanned paper archive at the SAO/NASA Astrophysics
Data System (ADS, adsabs.harvard.edu/cgi-bin/signup\_ocr) were exploited
whenever possible. It turned out that more often than not, these tables
either needed heavy reformatting or a complete rescanning and subsequent
OCR treatment. Before being accepted in the author's collection, all
OCR results are checked for their consistency with the original by
overlaying the OCR result printed on the same scale as the original,
which turned out to be the most efficient method to spot errors in the
OCR result. Frequently, errors in the original papers were also detected
and reported as notes to the resulting electronic table. These errors
show a certain lack of attention the referees pay to the data part of
papers. Ironically, it is usually the data part that remains valid for
many years after publication, while ideas about their interpretation may
change. The main bottleneck of making these tables available for catalog
browsers is the need for preparing metadata, which implies complete and
consistent column descriptions, byte per byte. These can in most cases
be taken from the text (column descriptions) contained in the papers,
although their preparation requires a person with minimum knowledge of
the research field.

\bigskip
{\large\bf 3~~~~~~~THE CATALOG COLLECTION AT CDS}

The CDS at Strasbourg maintains the most complete collection of
astronomical catalogs accessible for public downloads.  As of September
2008, the CDS collection offers catalogs from about 7700 publications,
growing now at over 500 items per year. In Figure~2, the solid line shows
the total number of items in the CDS collection as function of publication
year. (Note the difference in the meaning of the abscissae in Figures 1
and 2.) About 90 percent of these catalogs are also incorporated in the
VizieR catalog browser, which allows a search through all or a subset
of these catalogs around a user-specified sky position (dot-dashed line
in Figure~2).  The sudden increase of the slope of the continuous line
in Figure~2 near 1994 is due to the start of the electronic version of
the major astronomical journals and specifically due to an agreement
between CDS and the journal {\it Astronomy \& Astrophysics} (A\&A) to store its
data tables directly in the CDS archive. The 10 percent of CDS catalogs
not in VizieR are either still in the process of preparation of their
metadata for inclusion into VizieR or are unsuitable for cone searches
(e.g.\ for lack of absolute coordinates or suitable documentation).

The dashed line in Figure~2 shows the cumulative distribution of
publication years of catalogs in the author's catalog collection. It is
noteworthy that for papers published between 1982 and 1992, my collection
exceeds in number the catalogs available at CDS, while the overlap between
the two collections is small. This is due to my OCR activities on tables
from a period that immediately precedes the electronic journal era. But,
as I explain below, the more recent catalogs in my collection are by no
means a duplication of efforts at CDS since a large fraction of these
catalogs have still not entered the CDS archive. For the last few years
I have also monitored the catalogs listed as ``in preparation" in the CDS
archive (dotted line in Figure~2). The currently 300 such catalogs have
a median age since publication of nine years, compared to 2.5 years for
426 catalogs in 2003 (Andernach 2003) and five years for 400 of these in
2006 (Andernach 2006). While the decrease in number of these ``inprep"
catalogs may appear promising, it only implies that CDS has managed to
reduce the backlog by including preferentially the more recent ``inprep"
catalogs, but, as I show below, CDS appears to have stopped to list
more recent catalogs as ``inprep" when it was not clear when they would
be able to include these in the archive.

\begin{figure}[!t]
\hspace*{18mm}
\psfig{file=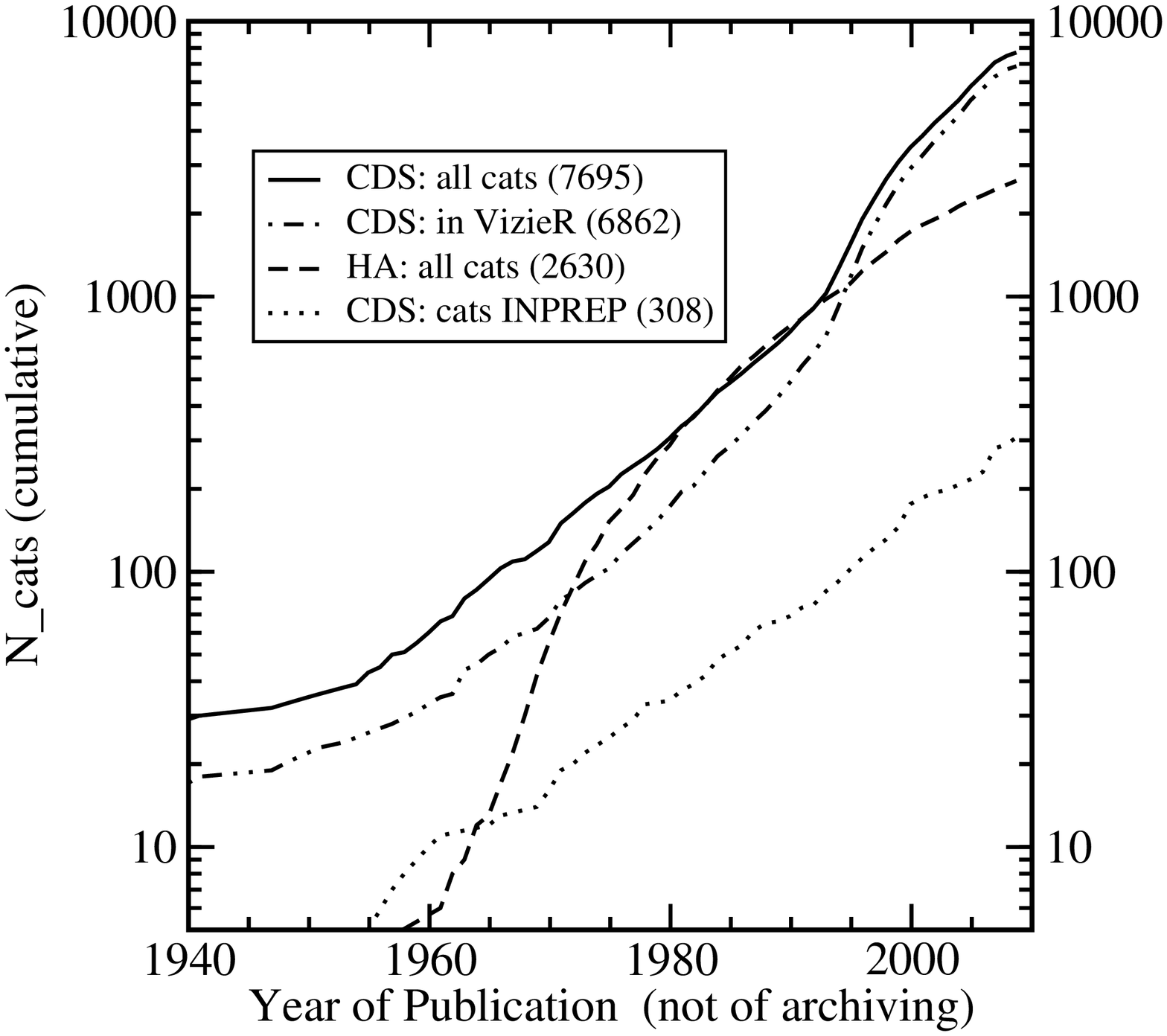,width=11cm,bbllx=572pt,bblly=4pt,bburx=83pt,bbury=493pt,angle=-90,clip=}
\caption{Comparison of the cumulative number of various catalog
collections, as function of publication year, as of Sept.~2008. The
continuous curve shows the full CDS collection, the dot-dashed curve
those in VizieR, the dashed curve those in the author's collection and
the dotted curve the catalogs listed as ``in preparation'' in the CDS
collection. Note the logarithmic scale of the ordinate.}
\end{figure}

\bigskip
{\large\bf 4~~~~~~~CATALOG ``BIOMETRICS'' AND DATA CENTER COVERAGE}

In Figures 3 and 4 (continuous lines) I have plotted the cumulative
size distribution of the radio- and non-radio catalogs in my
collection. Here ``size'' typically means the number of objects a catalog
deals with. Sometimes this is the number of flux density measurements,
e.g.\ if cataloged for various observing frequencies or epochs, and can
be much larger than the number of objects.  Since these plots are on a
double-logarithmic scale, it can be seen that the size distribution of
both types of catalogs closely follows a power law with similar slopes
of $-$0.69 and $-$0.61, respectively.  Such power laws are known as Zipf's or
Lotka's laws in biometrics. The dashed lines in Figures 3 and 4 indicate
the size distribution of those catalogs in my collection that are also in
the CDS collection. While it can be seen that the CDS collection becomes
incomplete for radio catalogs of fewer than about 10,000 records, it is
incomplete for the non-radio catalogs in my collection for all sizes. In
Figure~3, the dotted line gives the size distribution of those catalogs
in my collection that are also in the CATS collection, which become
incomplete at a much smaller size of about 1000 records. It would have
been interesting to create such plots for the entire collection of over
7000 CDS catalogs, but their sizes were not readily available to me.

Obviously the catalogs in all three collections have a typical
minimum size of roughly 50 records to be worth of being included in
the collection. The assumption that for all publications containing
astronomical data, regardless of their presence in catalog collections,
the power law for the number of objects holds down to very few records,
would lead, e.g.\ by extrapolating the continuous line in Figure~3, to an
estimate of $\sim$20,000 papers that have ever dealt with at least 

\begin{figure}[!t]
\hspace*{22mm}
\psfig{file=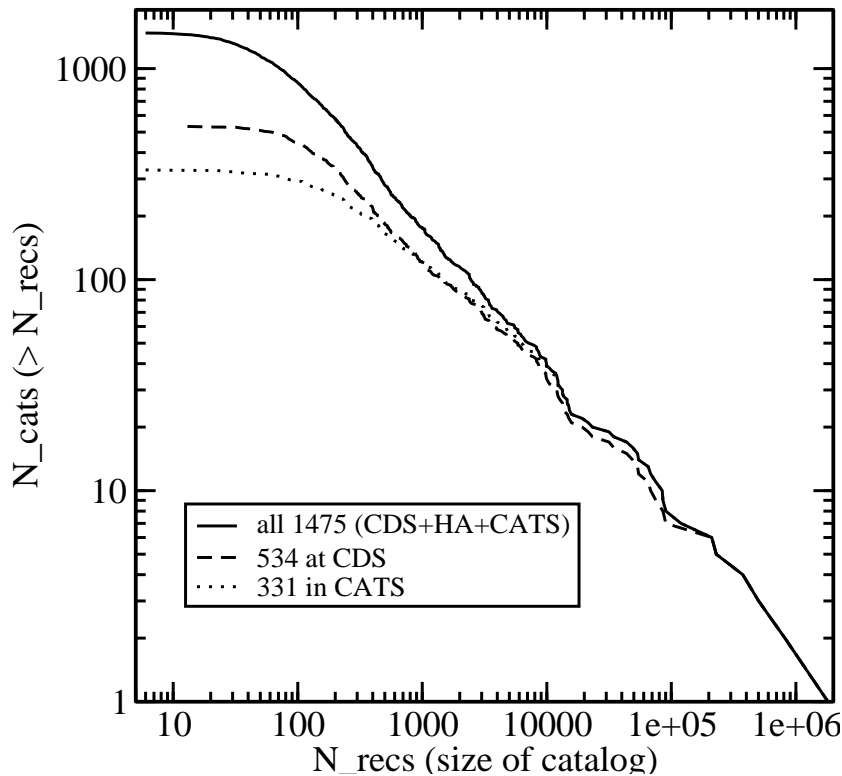,width=11cm,bbllx=567pt,bblly=16pt,bburx=84pt,bbury=526pt,angle=-90,clip=}
\caption{Cumulative size distribution of 1484 radio catalogs. For any
given size (in records) listed on the abscissa, the ordinate gives the
total number of catalogs up to that size (as of Sept.~2008). Continuous
line: author's collection plus those at CDS and CATS; dashed line:
subset of radio catalogs that are available at CDS; dotted line: subset
of radio catalogs that are available in CATS.}
\end{figure}

\begin{figure}[!h]
\hspace*{22mm}
\psfig{file=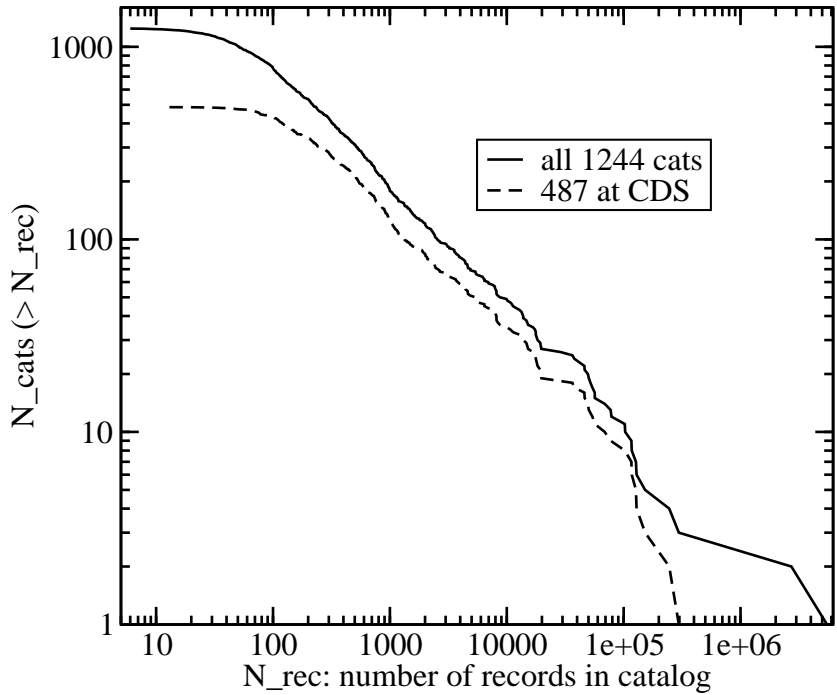,width=11cm,bbllx=570pt,bblly=15pt,bburx=86pt,bbury=586pt,angle=-90,clip=}
\caption{Cumulative size distribution of 1244 non-radio catalogs in the
author's collection as of Sept.~2008. Axes are as described in Figure
3. Continuous line: author's collection; dashed line: subset of the
latter catalogs that are available at CDS.}
\end{figure}

one radio
source. As the latter number seems reasonable, one may infer from the
turnover of the power law (below about 100 records) that the collections
become incomplete for smaller sizes. This is a natural ``collection bias''
since it is more work for less return to transfer into electronic form
the very many smaller catalogs, e.g.\ via OCR methods.

Figure~3 includes all radio source lists that exist in electronic form,
still excluding many dozens of published lists that have not (yet)
been recovered in electronic form. Thus, one may conclude that of all
existing electronic radio source catalogs, virtually all the ones larger
than 10,000 records are available from CDS, and all the ones larger than
1000 records are included in CATS, but for those larger than 100 records,
only 52 percent are in the CDS archive, and only 34 percent are included
in the CATS catalog browser.

From Figure~4, one sees that for those non-radio catalogs in the
author's collection, the CDS archive contains 69 percent of those
larger than 1000 records and 56 percent of those larger than 100
records. As opposed to Figure~3, however, Figure~4 only contains a
biased subsample of catalogs that were of some interest to the author,
and thus the completeness of the CDS archive cannot be assessed as a
whole on the basis of these data. Nevertheless, Figure~4 shows that of
the 774 catalogs with at least 100 records collected by the author, 344
(46 percent) are missing in the CDS archive. This percentage decreases
only slightly when a fraction of items in my collection, which may not
be suitable for inclusion in VizieR or archiving at CDS, e.g.\ for lack
of object coordinates or adequate documentation, are discarded.

\newpage
{\large\bf 5~~~~~~~LITERATURE COVERAGE OF VIZIER AND CATS}

Based on a systematic inspection of all major astronomical journals for
the period 1987 through 1993, I had found (Andernach, 1994) that there
were 374 articles that deal with at least 100 supposedly extragalactic
objects, and I showed that in 1994 the tables for only 21 percent of
these articles were available from the CDS archive. Today, CDS has 46
percent of these same 374 articles, showing that some of the published
data of the past are being recovered, albeit at a very slow pace. Table
1 gives an overview of the coverage of the CDS and CATS of those items
in my collection. One can see that overall only about 40 percent of my
collected items is covered.

{\bf Table~1.} For each catalog collection in the first column, I list the
number of items in my collection of radio and non-radio catalogs, followed
by the percentage in number of catalogs, the total number of records
of the catalogs and its percentage of the total. The status is as of
1-Sep-2008. HA stands for the author, and the third row refers to catalogs
which are in the author's, but not in the CDS or CATS collections. \\[-8mm]
{\small
\begin{verbatim}
                       RADIO source tables in HA coll.|  Non-radio tables in HA coll.
Collection              Ncats %cats   Nrecords  %recs | Ncats %-cats  Nrecords %-recs
------------------------------------------------------|------------------------------
VizieR (not INPREP)       534  36.0   5,238,510  91.3 |   487  39.1   2,462,060  21.7
CATS collection           343  23.1   5,418,785  94.6 |     8   0.7   2,800,896  24.7
HA collection ONLY        836  56.4     163,977   2.9 |   734  58.9   6,134,291  54.1
HA+CDS collections       1480 100.0   5,736,178 100.0 |  1244 100.0  11,340,943 100.0
\end{verbatim}
}

\medskip
Thus, for radio sources, CATS slightly ``beats'' VizieR in number of
objects, but VizieR offers more variety, i.e. more of the smaller source
lists. This is due to CDS's continuous activity of incorporating tables
of all sizes from the major electronic journals in astronomy. Given this
difference between the two collections, one needs to search in both to
obtain the most complete results. Unfortunately it is difficult to weed
out the duplications between the two collections. For both radio- and
non-radio catalogs, more than half of the items I collected is neither
in VizieR nor in CATS. For the vast majority of the latter catalogs,
metadata would have to be prepared from column descriptions given in the
paper text. For a small fraction of these catalogs (e.g.\ those lacking
absolute coordinates for the objects they contain), this would require
a further effort of inserting these coordinates, which could partly
be achieved with existing name resolvers of NED or SIMBAD. A still
smaller fraction of these catalogs, e.g.\ those that only contain derived
parameters, may be unsuitable for their inclusion into catalog browsers.

\bigskip
{\large\bf 6~~~~~~~THE ELECTRONIC AGE SINCE 2001: NOT ALL GOLD}

The above statistics may be distorted by the fact that catalogs published
in the pre-electronic age before about 1995 may be over-represented in
my collection and under-represented in the CDS collection. I report in
Table~2 the situation for tables with over 50 records in my collection,
both radio and non-radio, and published in electronic journals since 2001.
Only journals with at least five catalogs in my collection, and published
in this period, are included. Percentages in parentheses are for the
period from 1998 to 2006 as taken from Andernach (2006).

\medskip
{\bf Table~2.} Presence at CDS of journal tables in the author's collection,
published 2001 through mid-2008 and larger than 50 records. For the full
journal titles refer to the web page \linebreak[4]
cdsweb.u-strasbg.fr/simbad/refcode.html. For
each journal, I give the number of catalogs, the sum of their records,
the number and percentage by number of those available at CDS and (in
parentheses) the same percentage found for the period 1998 to 2006,
followed by the sum of their records, the percentage of records, and
(in parentheses) the percentage by records for the period 1998 to 2006. \\[-4ex]
{\small
\begin{verbatim}
           Journal  N_cats   N_recs   N_CDS  % (2006)  N_rec_CDS  % (2006)
          -----------------------------------------------------------------
           A&A       124     93,806      89  72  (62)     88,035  94  (78)
           AJ        151  3,720,917     104  69  (49)    962,266  26  (75)
           ARep        8     10,268       7  88  (54)     10,068  98  (16)
           Ap/Afz     12      3,838       1   8  ( 0)        122   3  ( 0)
           ApJ        65     86,228      30  46  (44)     21,461  25  (46)
           ApJS       68    579,256      46  59  (67)    499,543  86  (55)
           ChJAA       5      2,919       0      ( -)          0   0  ( -)
           MNRAS     152    847,607     101  66  (54)    568,413  67  (64)
           PASA        5     39,011       3  60  (17)     38,246  98  ( 1)
           Total     590  5,383,850     381  65  (54)  2,188,154  41  (70)
\end{verbatim}
}

Table~2 shows no general trend for bigger catalogs to be more likely
found in the CDS archive and only a small improvement between the
period from 1998 to 2006 and that from 2001 to 2008. However, the exact
percentages of the coverage by records of course depend on the inclusion
or not of a few very large catalogs (e.g.\ the Sloan Digital Sky Survey,
SDSS, www.sdss.org). While the above statistics are biased by my way
of selecting radio and extragalactic source tables, the large number
of articles they are based on still shows a significant fraction of
data missing from the data centers, which has not improved in recent
years. Moreover, the following problems are often faced with electronic
journal tables.

The journals ApJ and AJ offer most tables as separate files in ASCII
format, usually with good metadata provided by the authors, but often
the tables are not well aligned and mixed with HTML or Latex symbols;
the recent change of publisher of these journals from the University
of Chicago Press (USA) to the Institute of Physics (UK) has lowered the
percentage of well-formatted and documented tables.

In A\&A many tables are offered only in Latex or PS, but not in ASCII
format, nor do they come with metadata. Despite the 1994 agreement between
the editors of A\&A and CDS to offer A\&A data tables automatically at CDS,
the editors decide which tables flow to CDS, and for the sample in Table
2 the coverage is only 72 percent.

In the UK journal MNRAS, data tables are not usually offered in ASCII,
except for articles with supplementary material. The tables are mostly
offered in a separate window which does not offer to download them as
files, but only in cut-and-paste mode, which becomes difficult for long
tables. Metadata are not common. Many tables are not offered as separate
files at all and need to be recovered by pdf-to-text conversion. It is
often easier to recover these tables in Latex format from the public
preprint archive at {\tts www.arXiv.org}, but then there is no guarantee that
they are identical with the published version. The publisher of MNRAS
declines any responsibility for supplementary material such as data
tables and refers to individual authors in case of problems. This does
not show an attitude towards a long-term preservation of published data.

For other journals like ARep, BSAO, CHJAA, PASA, or RMxAA, no ASCII
tables are offered at all. They can only be recovered via a pdf-to-text
conversion.

A problem that applies to all journal tables not offered as a straight
ASCII file is that the cut-and-paste copying or pdf-to-text conversion
usually does not preserve table alignment and does not correctly interpret
special characters such as minus signs.

\bigskip
{\large\bf 7~~~~~~LITERATURE COVERAGE OF DATABASES}

The contents of the CDS catalog archive are easy to assess since the
URL ftp://vizier.u-strasbg.fr/cats/cats.all offers a full and always
up-to-date list (except for their subset in VizieR). However, to obtain
the bibliography and published data on an object, most astronomers
consult databases like SIMBAD, NED, or LEDA.  In the following, I try
to obtain an estimate of the fraction of tabular data covered by these
databases. It should be stressed that these databases are independent
of the catalog collections described above. Databases make use of
individual catalogs to link data to certain objects, but this requires
fairly sophisticated cross-identification procedures, which need to be
controlled by the database managers. Thus, it requires much more effort
to include the data of a certain catalog into a database, compared to
including a catalog into a searchable catalog browser. As a result,
a user may expect a fairly reliable set of data for a given object put
together in a database, while one may obtain more complete data using
cone searches around an object in catalog browsers if one is ready to
check which of the data actually correspond to the object in question.

As SIMBAD puts a weaker emphasis on extragalactic data and LEDA
concentrates on galaxies in the nearby Universe, they were not
studied here. The literature coverage of NED is offered at the URL
nedwww.ipac.caltech.edu/samples/NEDmdb.html and is expressed for each
reference with terms such as ``incomplete'', ``partial", ``complete", or
``entered as found". For 82 percent of the altogether 4637 references
listed in the above file, NED claims that the objects were completely
entered. For 5 percent of the references, NED explicitly admits
incompleteness or a mere lack of the electronic source catalog, and for 7
percent objects were only entered ``as needed'' or as references to them
were found in the literature.  For one percent of the references, only
the likely extragalactic objects were entered in NED. The disadvantage
of the NEDmdb list is that it is apparently not updated frequently
since in late September 2008 that list was of 27-Mar-2008. Moreover,
occasionally objects in NED are attached to bibliographic references
for which NED decided to create a specific nomenclature (acronym) for
an object. This latter problem does not persist for the kind of study
I describe in the following paragraph.

For a more quantitative study, I did a NED search ``by refcode'' for all
articles for which I had collected tables, so as to obtain the number
of NED objects per article, and compared this with the full catalog
size. This is not always reasonable, as some tables give several records
per astronomical objects, while other tables may not be of relevance to
NED (e.g.\ Galactic radio sources, etc.). Moreover, some of the largest
datasets (e.g.\ NVSS, or the various data releases of SDSS) have a
different refcode in NED than their publication refcode. Despite all
these limitations, my results are summarized in Table~3. Note that I
did not limit the catalogs to a minimum size, as NED should recognize
a refcode with any small number, if only extragalactic, objects. One
should also note that the presence of a certain catalog in NED does not
mean that the user is able to retrieve the entire catalog's data content
from NED. It basically means that NED has made a link between all catalog
objects (or a fraction thereof) and its bibliographic reference. Catalog
browsers generally offer the content of all catalog columns.

\medskip
{\bf Table~3.} NED coverage of data in catalogs of the author's collection,
as of Sept.~2008.  The second column gives the total number of catalogs
in each category in the author's collection, followed by number and
percentage by number of those catalogs for which at least half of the
objects are included in NED. The last two columns give the number and
percentage of catalogs in the author's collection which are not covered
by NED. \\[-3ex]
{\small
\begin{verbatim}
                                               at least 50%   |    no object
       Catalog Category        Number of      objects in NED  |   found in NED
                               HA tables      N-cats  %-cats  |   N-cats  %-cats
       -------------------------------------------------------|--------------------
       Radio source tables       1361          780   57%      |    480   35%
       Non-radio tables          1230          887   72%      |    225   18%
\end{verbatim}                                                   
}                                                                

\medskip
Given the above-mentioned limitations, these numbers are only indicative
but suggest a more significant lack of radio source data. Indeed, a few
of the largest recent radio surveys, such as WENSS and WISH, Miyun, 7C,
UTR, VLSS, and SUMSS, were not (or almost not) represented in NED. This
is where systems such as CATS (Verkhodanov, Trushkin, Andernach, \&
Chernenkov, 2008) complement, offering more data, albeit leaving the
cross-identification work to the user.

\bigskip
{\large\bf 8~~~~~~~CONCLUSIONS}

The teams of CDS and NED are doing their best to cope with the avalanche
of data, but human resources do not suffice to achieve completeness
above about the 50-percent level. Bigger catalogs are preferentially
incorporated, but it is the medium- and smaller-sized catalogs that
offer wavebands or observing epochs different from the bigger ones and
are thus important for a more complete multi-wavelength research. Based
on a systematic search in the major astronomical catalog browsers and
databases, for the presence of data from 2600 articles containing data
on radio sources and extragalactic objects, and for which I collected
the electronic data tables, I come to the following conclusions.

Less than 70 percent of currently published journal tables with more
than 50 entries become accessible through catalog browsers (VizieR,
CATS) routinely. We are definitely far from an ``automatic data flow"
from current electronic journals towards the data centers. This should
not be left to the willingness of the authors, who tend to be under
severe publication pressure and many lack the time to work on making
their tables available in an appropriate form.

The completeness of databases such as NED, SIMBAD, and LEDA is
more difficult to assess and quantify, but the present study reveals
significant shortages, at least for NED and especially for radio source
data.

Catalog browsers are easier to maintain or upgrade than databases such
as NED since they avoid the tedious task of cross-identifications,
but they can offer valuable data more rapidly to the interested (and
knowledgeable) user.

Current data centers are overloaded with the rate of data published
in journals. For the future Virtual Observatory, it is necessary to
dedicate more emphasis (a) to recover tables from the ``pre-electronic age"
before about 1995 and (b) to prepare adequate metadata for these tables,
e.g.\ those collected by the present author, to make them searchable.

We need more collaboration between authors, referees, journals, and the
data centers to make more data available immediately after publication
and safeguard them for the future.

The author is (and has always been) ready to provide his table collection
to data centers and database managers (who would only need to write
the metadata).  For recovery of further tables the ``heritage" of the
LEDA team, which is suffering from a lack of long-term support but has
scanned a vast amount of the pre-electronic literature on galaxies,
should be salvaged.

\bigskip
{\large\bf 9~~~~~~ACKNOWLEDGMENTS}

I am grateful to various students and secretaries who in the course
of time have helped in recovering or proofreading journal tables via
OCR. Many tables were recovered also from the OCR offered at ADS, albeit
with heavy reformatting and correction work. Thanks to F. Ochsenbein for
providing me with an up-to-date list of VizieR catalogs and to N. Aguilera
Navarrete for converting this paper into a Word document. I acknowledge
support from grant 81356 from CONACyT of Mexico and the hospitality of
the Emmy-Noether Research Group of T. Reiprich at AIfA, Univ. of Bonn,
Germany, where partial support from the Transregional Collaborative
Research Centre TRR33 ``The Dark Universe" was received. I am grateful to
an anonymous referee for comments that helped clarify parts of this paper.

\bigskip
{\large\bf 10~~~~~~REFERENCES}

Andernach, H. (1992) Steps Towards a Radio Source Data Base, in Heck, A. \&
Murtagh, F. (Eds.), {\it Astronomy from Large Data Bases - II}, ESO Conference \&
Workshop Proceedings, 43, 185--190, Garching, European Southern Observatory

Andernach, H., (1994), How Complete is the Electronic Archive at CDS
for information on Extragalactic Objects\,?, {\it Bull. Inf. CDS} ~45, 35--45
({\tts cdsweb.u-strasbg.fr/Bull/45/chapter1\_2\_3.html})

Andernach, H. (2003) Accessibility and Exploitation of ``Published''
Extragalactic Data: Public and Private Efforts, electronic
version of an invited talk given at the Guillermo-Haro Workshop
on {\it AGN surveys}, July 2003 at INAOE, Mexico 
~({\tts www.inaoep.mx/survey03/heinz.html})

Andernach, H. (2006) How Accessible are Extragalactic and Radio Data
from Journal Tables\,?, Poster presented at XXVI IAU General Assembly,
Special Session~6 ~{\it Astronomical Data Management}, Prague, August 22, 2006
~({\tts adsabs.harvard.edu/abs/2006IAUSS...6E...7A})

Harris, D. E., Stern Grant, C. P., Andernach, H., (1995) The EINSTEIN
On-Line Service, in Shaw, R.A., Payne, H.E. \& Hayes, J.J.E. (Eds.),
{\it Astronomical Data Analysis Software and Systems IV}, San Francisco:
Astronomical Society of the Pacific Conference Series 77, 48--51

Mazzarella, J. M., \& The NED Team (2007) NED for a New Era. in Shaw,
R.A., Hill, F. \& Bell, D.J. (Eds.), {\it Astronomical Data Analysis Software
and Systems~XVI}, San Francisco: Astronomical  Society of the Pacific
Conference Series 376, 153--162

Ochsenbein, F., Bauer P., \& Marcout, J. (2000) The VizieR database of
astronomical catalogues. {\it Astron.\ \& Astrophys. Suppl.\ Ser.} 143, 23--32

Paturel, G., Andernach, H., Bottinelli, L., Di Nella, H., Durand, N.,
Garnier, R., Gouguenheim, L., Lanoix, P., Marthinet, M.-C., Petit,
C., Rousseau,  J., Theureau, G., \& Vauglin, I. (1997) Extragalactic
database. VII. Reduction of astrophysical parameters. {\it Astron.\ \&
Astrophys.\ Suppl.\ Ser.} 124, 109--122

Verkhodanov, O. V., Trushkin, S. A., Andernach, H., Chernenkov,
V. N. (1997) The CATS database to operate with astrophysical catalogs,
in Hunt, G. \& Payne, H. E. (Eds.), {\it Astronomical Data Analysis Software
and Systems~VI}, San Francisco: Astronomical Society of the Pacific
Conference Series 125, 322--325

Verkhodanov, O. V., Trushkin, S. A., Andernach, H., Chernenkov,
V. N. (2009) The CATS database: an astrophysical research tool.
{\it Data Science Journal}, Vol.\ 8, in press \\
({\tts http://dsj.codataweb.org})

Wenger, M., Ochsenbein, F., Egret, D., Dubois, P., Bonnarel, F., Borde,
S., Genova, F., Jasniewicz, G., Lalo\"e, S., Lesteven, S., \& Monier,
R. (2000) The SIMBAD astronomical database. The CDS reference database
for astronomical objects. {\it Astron.\ \& Astrophys.\ Suppl.\ Ser.} 143, 9--22

\end{document}